\RequirePackage{fix-cm}
\documentclass[smallextended]{svjour3}       
\smartqed  

\usepackage[dvips]{graphics,epsfig}
\usepackage{multirow}
\usepackage{color}
\usepackage{ulem}
\usepackage{amsmath}

\begin{document}

\title{Comprehensive cluster validity Index based on structural simplicity}



\author{Anri Mutoh         \and
        Masamichi Wada     \and
        Kou Amano
}


\institute{Anri Mutoh \at
             National Statistics Center, Wakamatsu-cho, Shinjuku-ku, Tokyo 162-8668, Japan\\
              Tel.: +81-80-6445-4368\\
              \email{anry.12.25@gmail.co.jp}\\
              The document is the author’s responsibility and does not reflect the institute.
           \and
           Masamichi Wada \at
             Graduate School of Library, Information and Media Studies, University of Tsukuba, 1-2 Kasuga, Tsukuba-shi, Ibaraki 305-8550, Japan \\
            \and
           Kou Amano \at
              Research and Services Division of Materials Data and Integrated System, National Institute for Materials Science, 1-2-1 Sengen, Tsukuba, Ibaraki 305-0047, Japan
}

\date{Received: date / Accepted: date}

\maketitle

\begin{abstract}\mbox{}
Nonhierarchical clustering depending on unsupervised algorithms may not retrieve the optimal partition of datasets.
Determining if clusters fit ``natural partitions'' can be achieved using cluster validity indices (CVIs). 
Most existing CVIs consider criteria such as cohesion, separation, and their equivalents. 
However, these binary relations may provide neither the optimal measure of partition suitability nor reference values corresponding to the worst partition. 
Moreover, previous CVI studies have been mostly focused on fitting correct partitions according to researchers’ a priori assumptions.
In contrast, we investigated desirable properties of CVIs, namely, scale--shift transform invariance, optimal clustering, and unbiased clustering with representing the worst partition. 
Then, we conducted experiments to evaluate whether existing CVIs fulfill these properties. 
As none of these CVIs fulfilled the desired properties, we propose the simplicity index, which measures the simplicity of tree structures in clusters. 
The simplicity index is the unique index invariant to the ``correct rate'' and provides both a reference indicating the most complex partition and the best value indicating the simplest one.


\end{abstract}

\section*{Introduction}
Clustering aims to automatically classify data points and is widely used in various fields such as biology, humanities, and informatics \cite{Mirkin2012}.
A clustering algorithm should classify (group) data according to their underlying structure \cite{Luxburg2012}, such that similar members are assigned to the same cluster, and diverse members belong to different clusters \cite{Sneath1973}.
\par
Several automatic clustering approaches are available and belong to either of two schemes: hierarchical and partitional (nonhierarchical) \cite{jain1999}. 
In both schemes, clustering algorithms perform grouping according to some criteria, and thus they may not retrieve suitable partitions.
Cluster validity indices (CVIs) measure the fitness of clustering results to ``natural partitions'' \cite{Halkidi2001} \cite{Jain2010} \cite{Arbelaitz2013}.
Most existing CVIs are based on similarity criteria, such as cohesion and separation, but do not emphasize the global structure of clusters or their underlying properties.
\par
In this paper, we establish desirable properties for CVIs and propose a novel nonhierarchical CVI based on simplicity of the tree structure in clusters.


\section*{History of classification and clustering}
Clustering originally derived from taxonomy and adopted its name after its development was mature. 
Classification has more ancient references than taxonomy with its mentions dating before Christ. 
In fact, the origin of classification can be found in ancient libraries \cite{casson2002}, and the notable philosophers Plato and Aristotle tried to understand the structure of all things through classification. 
For instance, Aristotle proposed category theory and classified a wide variety of things and phenomena \cite{Ammonius1991}. 
This was likely the first classification specialized in natural sciences, and Aristotle's category theory influenced the succeeding taxonomy. 
However, a formal methodology for classification was not established, and a unified discussion about classification from that period has not been found.
\par
In the 18th century, classification was studied in the form of taxonomy in biology. Linnaeus classified animals, plants, and minerals using a unified system and introduced his taxonomy in the publication titled `Systema Naturae' in 1735 \cite{anderson2009}. 
Linnaean taxonomy has a hierarchical structure composed of classification units such as class, order, and genus, thus establishing the basis for modern taxonomy. However, it lacks the perspective of genetic relationships but only considers morphology similarity.
\par
In 1840, Hitchcock introduced a diagram like a phylogenetic tree in his book \cite{archibald2009}. Although he did not claim the concept of biological evolution, it seems that he had the concept of hierarchy by genetic relationships.
Then, in the mid-20th century, several scientific methods were established, and both the evolution and genetic theories developed.
\par

Then, taxonomy was criticized for setting classification through arbitrary morphological properties, which did not conform a scientific method.
Therefore, in 1950, Hennig proposed cladistics, a more robust method that compares many morphological properties and performs classification according to the smallest number of branches (this method is equivalent to maximum parsimony) \cite{schmitt2003}. Afterwards, research on classification became increasingly quantitative. 
Pearson (1926) \cite{pearson1926} proposed a method to classify races based on the similarity of groups, paving the way for automatic classification using computers.
Research on automatic classification, which can be called clustering, was simultaneously initiated at various places but using different expressions. 
In 1958, Steinhaus proposed the $k$-means algorithm \cite{hans2008}, and Cox (1957) \cite{cox1957} proposed a method that became the prototype of clustering using the sum of squares as criterion. 
\par

Around the 1960s, research on clustering validation began, with MacQueen studying the validation of the $k$-means algorithm in 1967. 
Currently, three criteria are available for clustering validation \cite{Theodoridis2003}. 
First, external criteria that evaluate the similarity of a partition to a priori clusters. These criteria assume that there exists a correct partition.
Second, internal criteria that use measures from the data themselves (e.g., proximity matrix). 
These two types of criteria often involve statistical tests and Monte Carlo methods. 
Finally, relative criteria that decide the optimal partition by comparing various candidate partitions. 
The CVIs are measures that allow comparing various partitions, and thus can be considered relative criteria.
\par
The major CVI studies are Dunn's index\cite{Dunn1973}, Calinski-Harabasz index \cite{Calinski1974}, Silhouette index \cite{Rousseeuw1987}, and Score Function \cite{Saitta2007}. 
In addition,  after 2010 there are several review articles of CVI, Albelaittz, et.al(2013) conducts a review of 30 indices.


\section*{State of the art in CVI}
CVIs are more recent than classification and clustering, and quantitative clustering has been developed since the 20th century. 
Most existing CVIs only consider data points in clusters. 
For instance, the distance among either members within a cluster to define cohesion or clusters to define separation \cite{Gurrutxaga2010} is commonly used in CVIs.
Cohesion represents the closeness of elements in a cluster and is usually given by the mean distance between the cluster centroid and members.
Separation represents the distinctness between two clusters and is usually given by the distance between centroids of the clusters.
Hence, smaller distance in cohesion and larger distance in separation indicate better clustering.
\par
The difference or ratio between cohesion and separation can bias clustering. 
For instance, if only one cluster exists, the distance between clusters becomes zero, and the distance among cluster members becomes maximum.
On the other hand, if as many clusters as data points exist, the distance between clusters becomes maximum, and the distance among cluster members becomes zero.
Therefore, CVIs considering cohesion and separation are biased towards maximizing the number of clusters.

\par
Furthermore, most CVIs don't have a best value. In this context, the best value is the maximum (or minimum) value defined by the formula for the CVI. Infinity is the largest value, but not the best value.
Cluster validation using relative criteria compares different partitions, and the optimal partition is usually selected from the CVI according to the number of clusters. 
In this curve, only notable local changes or equilibria appear \cite{Halkidi2001}, and hence different CVIs can retrieve varying optimal partitions.
\par
CVIs have been reviewed in few articles \cite{Milligan1985} \cite{Arbelaitz2013}. Most studies have compared few CVIs with data in special fields and focused on increasing the correct rate for datasets to conform to a suitable a priori partition. However, they have mostly neglected the structure and desirable properties of clusters when developing CVIs. 


\section*{Desirable properties for CVI}

We propose properties based on existing research on CVI aiming optimal clustering and to avoid clustering bias. We consider these properties as desirable to retrieve a suitable clustering.

\begin{enumerate}
        \item Transform invariance: Clustering should not depend on scale or shift transformations (i.e., linear transformations). This way, clustering can be evaluated by absolute measures, besides relative structural measures. Such invariance has been overlooked in the existing CVIs.
        \item Optimal clustering: The best value corresponding to the most suitable partition should be global. 
        For instance, grouped into a single cluster is the most suitable partition for the same data points, then grouped it into two clusters should get values worse.       
       \item Unbiased clustering: The number of clusters should impose no bias on the partition. This bias can appear as a preference for either the minimum or maximum number of clusters. 
\end{enumerate}

\section*{Evaluation of existing CVIs}

We evaluated the 30 existing CVIs presented in the survey by Arbelaitz et al. (2013) \cite{Arbelaitz2013} regarding the three above-mentioned properties, namely, transform invariance, optimal clustering, and unbiased clustering. The datasets and methods used in this evaluation are described in Table \ref{tab:1}.

\par

\begin{table}[!h]
\caption{Datasets and evaluation methods to evaluate desirable CVI properties} 
\label{tab:1}       

\begin{tabular}{cc}
\hline\noalign{\smallskip}
Evaluation method& Datasets\\
\noalign{\smallskip}\hline\noalign{\smallskip}
\multicolumn{2}{l}{\bf{Transform invariance}}\\
\hline

\begin{tabular}{l}
\\
$1)\;Equal(CVI(X_2),\,CVI(a\!\ast\!X_2))$\\
$2)\;Equal(CVI(X_2),\,CVI(X_2+b))$\\

\\
(for $a$ and $b$ scalars)\\
Calculate the scale transformation by\\
scalar multiplication of the data points\\
or constant shifting along the\\
direction of each basis. Then check\\
whether the corresponding CVIs differ\\
\\
Property flags:\\
$S\cdots$ true for the evaluations 1) and 2)\\
$s\cdots$ true for the evaluation 1) or 2)\\

\end{tabular}

&

\begin{tabular}{l}
(Two types of arguments:)\\
(Short dataset)\\
$X_2^S:= \textcircled{1}\, |\,\textcircled{2}\textcircled{3}$\\
(Long dataset)\\
$X_2^L:= \textcircled{1}\textcircled{2}\textcircled{3} 
\, |\,\textcircled{4}\textcircled{5}\textcircled{6}\textcircled{7}\textcircled{8}\textcircled{9}$\\
\end{tabular}\\

\hline\noalign{\smallskip}
\multicolumn{2}{l}{\bf{Optimal clustering}}\\
\hline

\begin{tabular}{l}
\\
$3)\;IsBestValue(CVI(Y_1))$\\
$4)\;IsBetter(CVI(Y_1),(than)\, CVI(Y_2))$\\
\\
When same data points all belong to\\
the same cluster, check whether CVI\\
reaches the best value. \\
When data points belong to\\
a different clusters, check whether\\
CVI value becomes worse. \\
\\
Property flags:\\
$B\cdots$ true for the evaluations 3) and 4)\\
$b\cdots$ true for the evaluation 3)\\
\end{tabular}

&

\begin{tabular}{l}
(Short dataset)\\
$Y_1^S:= \textcircled{1}\textcircled{1}\textcircled{1}$\\
$Y_2^S:= \textcircled{1}\textcircled{1}\, |\,\textcircled{1}$\\
(Long dataset)\\
$Y_1^L:= \textcircled{1}\textcircled{1}\textcircled{1}
\textcircled{1}\textcircled{1}\textcircled{1}$\\
$Y_2^L:= \textcircled{1}\textcircled{1}\textcircled{1}\,
|\,\textcircled{1}\textcircled{1}\textcircled{1}$\\
\end{tabular}\\

\hline\noalign{\smallskip}
\multicolumn{2}{l}{\bf{Unbiased clustering}}\\
\hline

\begin{tabular}{l}
\\$5)\;Equal(\;IsBaseLine(CVI(X_1)),$\\
\ \ \ \ \ \ \ \ \ \ \ \ \ \ \ \ \ \ $IsBaseLine(CVI(X_{3,9})\;)$\\
\\
When each cluster corresponds to\\
each data point, and\\
when only one cluster exists, \\
evaluate CVI has the base line\\
which corresponds to the most\\
complex state.\\
\\
Property flags:\\
$C\cdots$ true for the evaluation 5)\\
\end{tabular}

&

\begin{tabular}{l}
(Short dataset)\\
$X_1^S:= \textcircled{1}\textcircled{2}\textcircled{3}$\\
$X_3^S:= \textcircled{1}\, |\,\textcircled{2}\, |\,\textcircled{3}$\\
(Long dataset)\\
$X_1^L:=  \textcircled{1}\textcircled{2}\textcircled{3}
\textcircled{4}\textcircled{5}\textcircled{6}
\textcircled{7}\textcircled{8}\textcircled{9}$\\

$X_9^L:= \textcircled{1}\,|\textcircled{2}\,|\textcircled{3}\,
 |\textcircled{4}\,|\textcircled{5}\,|\textcircled{6}\,
 |\textcircled{7}\,|\textcircled{8}\,|\textcircled{9}$\\
\end{tabular}\\

\noalign{\smallskip}\hline

\multicolumn{2}{r}{
\begin{tabular}{r}
$X=\{\textcircled{1}\!:\!(0,\!0,\!1),\,\textcircled{2}\!:\!(0,\!1,\!0),\,\textcircled{3}\!:\!(1,\!0,\!0)\}$,
$Y=\{\textcircled{1}\!:\!(0,\!0,\!1),\,\textcircled{1}\!:\!(0,\!0,\!1),\,\textcircled{1}\!:\!(0,\!0,\!1)\}$\\

$|\,\!:\!partition$ (
$\textcircled{4}\!:\!:\!(0,\!0,\!2)$  $\textcircled{5}\!:\!:\!(0,\!2,\!0)$ $\textcircled{6}\!:\!:\!(2,\!0,\!0)$
$\textcircled{7}\!:\!:\!(0,\!0,\!3)$  $\textcircled{8}\!:\!:\!(0,\!3,\!0)$ $\textcircled{9}\!:\!:\!(3,\!0,\!0)$
)\\

When evaluating optimal clustering and bias, the denominator can\\
become zero. We interpret it as undefined state (no value).
\end{tabular}
}\\
\end{tabular}

\end{table}


\subsection*{Datasets}

We synthesized three-dimensional datasets to evaluate the CVIs according to the three desirable properties. We considered two types of datasets, namely, short and long. 
The short dataset consists of three data points, and the long datasets consist of six to nine data points. 


The long datasets are necessary when determining the point symmetry-based CVIs and graph theory-based CVIs collected by Arbelaitz et al. (2013), where the former CVIs are defined by a point symmetry distance \cite{Bandyopadhyay2008}, which is zero when fewer than three data points per cluster are available. 
\par
In addition, graph theory-based CVIs have the same expression as existing CVIs but use subgraphs of the cluster structure as data \cite{Pal1997}. 
The subgraphs can establish minimum spanning tree (MST), relative neighborhood graph (RNG), or Gabriel graph (GG). 
Hence, at least two points per cluster in the MST and three points in the RNG and GG are necessary, otherwise the data points of the subgraph will be those of the complete graph, thus being equivalent to a CVI not based on graph theory. Hence, we constructed the short and long datasets to prevent these situations. Short and long datasets are denoted in Table \ref{tab:1} by superscripts (S) and (L), respectively, and the number of clusters (1, 2, 3, 9) is indicated in the subscript : $X_{number\, of\, clusters}^{dataset\ type}$ .\\

\subsection*{Evaluation methods of the three properties}
We applied evaluation methods to verify the three properties in the clustering methods by introducing property flags indicating the fulfillment of each property (Table \ref{tab:1}).

For transform invariance, datasets were compared regarding homogeneity (scale) and additivity (shift). For optimal clustering, we considered the situation where data points at equal coordinates belong to the same cluster as one of the optimal partition corresponding to the best CVI value. For instance, dataset (\textcircled{1}$(0,0,1), $\textcircled{1}$(0,0,1), $\textcircled{1}$(0,0,1)$) is grouped into a single cluster, and (\textcircled{1}\textcircled{1}\textcircled{1}) may be optimal. 
Therefore, datasets consisting of the same coordinates were examined and partitioned. For unbiased clustering, we considered as key feature the presence of a reference value, a base line, which represents one of the worst CVI value corresponding to the most complex tree structure in the clusters. Since the base line is only a guide, we could retrieve it.
We define two states for the most complex structure: one cluster per data point, and a single cluster grouping all the data points (i.e., no partition). 
Therefore, datasets should be partitioned between these two extreme cases.
\par

\subsection*{Results}
The evaluation results are listed in Table \ref{tab:2},
where the CVI names are abbreviated, and the arrows indicate the tendency of improving values. 
Specifically, the upward arrow indicates that higher values are better, whereas the downward arrow indicates that lower values are better. 
The results marked with `*' have been retrieved from their original sources. 
\par

\begin{table}[htbp]
\centering
\caption{Evaluation results of the 30 CVIs discussed in Arbelaitz et al. (2013)}
\label{tab:2}       
\begin{tabular}{rcc}
\hline\noalign{\smallskip}
Evaluated CVI & Dataset & Fulfilled property\\
\noalign{\smallskip}\hline\noalign{\smallskip}
$CH(\uparrow)$ &$Short$& $S$\\
$Sil(\uparrow)$ &$Short$& $S\,b$*\\
$SF(\uparrow)$ &$Short$& $s\,b$\\
$Gamma(\downarrow)$ &$Short$& $s$\\
$C(\downarrow)$ &$Short$& $S$\\
$Davis\mathchar`-Boulding(\downarrow)$ &$Short$& $S$\\
$SDbw(\downarrow)$ &$Short$& $S\,B$\\
$CS(\downarrow)$ &$Short$& $S$\\
$DBv(\downarrow)$ &$Short$& $S\,b$\\
$COP(\downarrow)$ &$Short$& $S\,b$\\
$Negentropy(\downarrow)$ &$Short$& $S\,B$\\
$SV(\uparrow)$ &$Short$& $S$\\
$OS(\uparrow)$ &$Short$& $s$\\
$Dunn(\uparrow)$ & $Short$& $S$\\

$$&\\
\multicolumn{1}{c}{ \bf generalized Dunn }\\
$gD31(\uparrow)$ &$Short$& $S$\\
$gD41(\uparrow)$ &$Short$& $S$\\
$gD51(\uparrow)$ &$Short$& $S$\\
$gD43(\uparrow)$ &$Short$& $S$\\
$gD53(\uparrow)$ &$Short$& $S$\\

$$&\\
\multicolumn{1}{c}{ \bf {point-symmetry based} }\\%
$Sym(\uparrow)$ & $Long$& $S$\\
$SymDB(\downarrow)$ & $Long$& $S\,b$\\
$SymD(\uparrow)$ & $Long$& $S$\\
$Sym33(\uparrow)$ & $Long$& $S$\\

$$&\\
\multicolumn{1}{c}{ \bf {graph theory based} }\\
$Dunn\_MST(\uparrow)$ & $Long$& $S$\\
$Dunn\_RNG(\uparrow)$ & $Long$& $S$\\
$Dunn\_GG(\uparrow)$ & $Long$& $S$\\
$DB\_MST(\downarrow)$ & $Long$& $S$\\
$DB\_RNG(\downarrow)$ & $Long$& $S$\\
$DB\_GG(\downarrow)$ & $Long$& $S$\\

\noalign{\smallskip}\hline
\end{tabular}
\begin{flushright}
* Retrieved from \cite{Rousseeuw1987}\\
\end{flushright}

\end{table}


Clearly, no CVI fulfills all the desirable properties (scale--shift invariance, optimal clustering, and unbiased clustering), but many CVIs are scale--shift invariant. 
On the other hand, most CVIs have either an undetermined best value or a preference towards more (or fewer) clusters.
\par

\section*{Proposed CVI: simplicity index}
\subsection*{Overview}

Considering previous research and our evaluation of existing CVIs, we found that CVIs considering binary relations (i.e., cohesion and separation) as criteria can be biased and present an ambiguous best value.
Hence, other criteria besides cohesion and separation should be considered. 
\par
We propose the simplicity index ($SI$) as a CVI based on the concept of cluster simplicity, which allows to determine the underlying ``natural structure'' of data. By using simplicity as criterion, we can realize a CVI that is independent of the relation cohesion--separation, thus enabling compliance with the desirable properties.

\par
\begin{figure}
\begin{center}
	\epsfig{file=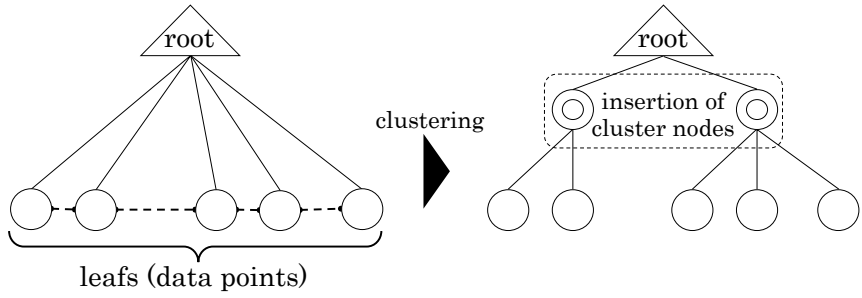,height=50mm}
	\caption{Structure of cluster tree to represent the clustering process. The leaves  correspond to the data points.  Nonhierarchical clustering is equivalent to inserting nodes into the cluster tree.}
	\label{fig:1}
\end{center}
\end{figure}

First, we consider the structure of cluster tree, whose features are listed below and depicted in Fig. \ref{fig:1}.

\begin{enumerate}
        \item The clusters are structured as a tree with a root and leaves being the data points.
        \item The distance between leaves can be defined.      
        \item A leaf has a path to the root.
        \item Non-hierarchical clustering inserts cluster nodes into a path from a leaf to the root just once.
\end{enumerate}

Then, we define the simplicity of a nonhierarchical cluster structure based on three considerations. Simplicity increases whenever:

\begin{enumerate}
        \item The number of clusters decreases. Adding clusters is equivalent to inserting nodes into a phylogenetic tree. However, this may not apply when the number of clusters drastically decreases.
        \item The radius of each cluster decreases. The radius indicates the mean distance among cluster members.
        \item The number of members in each cluster decreases.
\end{enumerate}

\subsection*{Definition}
The $SI$ is defined by the product of the three above-mentioned considerations. Equal values of $SI$ indicate the same degree of simplicity. 
If the number of members in each cluster is equal, a smaller radius is simpler. Conversely, if the radius of each cluster is equal, a smaller number of members is simpler. Thus, we define the $SI$ as follows:

\begin{equation} \label{eq:SI}
\begin{split}
SI = k \left(\prod_{n=1}^{k}{c_n^{\frac{r_n}{R}}}\right)^\frac{1}{k} \\
\frac{r_n}{R}=
  \begin{cases}
    0 & (R=0)\\
    \frac{r_n}{R} & (otherwise)
  \end{cases} 
\end{split}
\end{equation}

where $k$ is the number of clusters, $c_n$ is the number of members in the $n$-th cluster, $r_n$ is the radius of cluster $n$, and $R$ is the radius of the whole dataset. 
Whenever the denominator becomes 0, the numerator also becomes 0, so that the exponent part becomes 0 in order to prevent divergence.
\par

CVI $SI$ satisfies the desirable properties, namely, transform invariance, optimal clustering, and unbiased clustering. 
The best value of $SI$ is $1$, and the smaller the value, the better. When either there is one cluster per data point or a single cluster groups all the data points, $SI$ retrieves the number of data points.

Further, we propose a validation of $SI$ through distance matrix

\begin{equation} \label{eq:SI2}
\begin{split}
SI = k \left(\prod_{n=1}^{k}{c_n^{\frac{m_n}{M}}}\right)^\frac{1}{k} \\
\frac{m_n}{M}=
  \begin{cases}
    0 & (M=0)\\
    \frac{m_n}{M} & (otherwise)
  \end{cases}
\end{split}
\end{equation}

where $m_n$ is the mean of the pairwise distances among members of cluster $n$, and $M$ is the mean of the pairwise distances from the complete dataset.

\subsection*{Application to hierarchical structure}

\begin{figure}
\begin{center}
        \epsfig{file=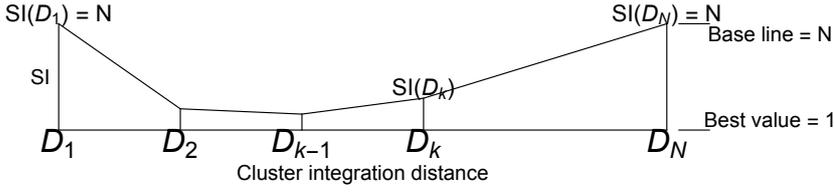,height=24mm}
        \caption{Evaluation of hierarchical clustering using SI. $SI_h$ evaluates the ratio of area $SI(D_1)$--$SI(D_k)$-- ... --$SI(D_N)$ and areas $SI(D_1)$--$D_1$--$D_N$--$SI(D_N)$}
        \label{fig:2}
\end{center}
\end{figure}

We applied the proposed $SI$ to a hierarchical cluster structure by determining it on each level of the hierarchical tree.
This way, we generated a plot of $SI$ according to the integrated distance for evaluating the hierarchical cluster tree as follows.
Assume $SI(D_k)$ is the evaluation of $SI$ for integrated distance $D$ at level $k$:

\begin{equation} \label{HSI}
SI_h = \frac{\sum_{k=2}^{N}{(SI(D_k) + SI(D_{k-1})) (D_{k} - D_{k-1}) / 2}}{(N-1) (D_N -D_1)} \\
; D_k \ge D_{k-1} .
\end{equation}
indicate evaluation of the complete tree structure (Fig.\ref{fig:2}).
allowing the evaluation of the complete tree structure, as shown in Fig. \ref{fig:2}.

On the other hand, we can consider the robustness of the generated curve.
The initial point corresponds to the cluster level where the number of clusters is the same as that of data points, and the final point corresponds to all data points being grouped into a single cluster.
If the hierarchical cluster structure is generated stably, the value gradually decreases reaching a minimum (optimal partition) and then gradually increases back to $N$, the number of data points.

\section*{Conclusion}
We considered that most existing CVIs only account for cluster cohesion and separation without emphasizing tree structures for clustering. 
Moreover, no study is available on the desirable properties for CVIs, but only surveys regarding the correct rate of CVIs have been developed.
Therefore, we propose three desirable properties that CVIs should possess and a CVI focused on the simplicity of tree clustering. The desirable properties are scale--shift invariance, optimal clustering, and unbiased clustering.

We evaluated 30 existing CVIs regarding these three properties on synthetic datasets and found that no CVI satisfies all of them. In fact, most CVIs do not have a best value and present bias. 
On the other hand, the proposed CVI $SI$ considers the simplicity of tree structures, the number of members in each cluster, and the relation between dataset and cluster radii, thus enabling the existence of best and reference values. Moreover, the proposed SI satisfies the three desirable properties for CVIs.


\begin{acknowledgements}
The authors would like to show our greatest appreciation to M. Sugimoto for collaboration in the early stages of this work. We would like to thank Pablo of Editage (www.editage.com) for English language editing.
\end{acknowledgements}




\end{document}